# Mapping The Best Practices of XP and Project Management: Well defined approach for Project Manager

Muhammad Javed[1], Bashir Ahmad[1], Shahid Hussain[2], Shakeel Ahmad[1]
[1]Institute of Computing and Information Technology Gomal University, D.I.Khan, Pakistan
[2] Namal University, Mianwali, Pakistan

Abstract--- Software engineering is one of the most recent additions in various disciplines of system engineering. It has emerged as a key obedience of system engineering in a quick succession of time. Various Software Engineering approaches are followed in order to produce comprehensive software solutions of affordable cost with reasonable delivery timeframe with less uncertainty. All these objectives are only satisfied when project's status is properly monitored and controlled; eXtreme Programming (XP) uses the best practices of AGILE methodology and helps in development of small size software very sharply. In this paper, authors proposed that via XP, high quality software with less uncertainty and under estimated cost can be developed due to proper monitoring and controlling of project. Moreover, authors give guidelines that how activities of project management can be embedded into development life cycle of XP to enhance the quality of software products and reduce the uncertainty.

——————————   ◆   ——————————

## 1 INTRODUCTION

Project management is considered as important component of certain domains including Information Technology (IT) and its main emphasis on the infrastructure of concern domain [5]. Improvement in success of project mainly depends on the proper management. Vague system understanding and improper documentation are the basic reasons for failure of any project which ultimately yields almost negligible productivity. Proper project planning is directly proportional to end result with a ratio of 1 to 4, as highlighted by "Margo20/80 theory"[6]. This shows project's productivity is directly related with planning process which is considered as primary activity of project management. Moreover, good project management leads to develop the high quality and less cost software. One of the key implementation of AGILE methodology is eXtreme Programming. XP is a collection of all the rules, practices and routines which have produced significant results in the past. [4]. A project being carried out using XP methodology is initiated by "Planning Game". This phase involves the interaction of key stakeholders i.e. XP practitioners and customers and end up with a list of functionalities to be developed. The Planning game is the only best practice which shows the activities of project manager of XP project while other best practices don't represent it. In this paper authors proposed a strategy to map the best practices of project management and XP process model. This strategy will help the project manager of XP to monitor and control all the activities during development life cycle. Moreover this strategy will lead to develop high quality software with less uncertainty and underestimated cost [8].

## 2 PROJECT MANAGEMENT

One of the critical process to observe during the software development project is "How Project Management is being carried out". A well managed Project Management produces positive outputs and helps to represent the sequence of activities which are performed by project manager. The challenging activities for a project manager are successful planning, controlling, coordinating, risk management and changing scope[1, 2]. All projects are developed with uncertain economy and increased pressures to derive optimal value. Planning is a time consuming and a laborious task in traditional/orthodox software development methodologies. While planning in XP provides milestones which are sorted by highest priority and are delivered at regular intervals [6]. The identification and delivery of iterative tasks are a challenge for project management. There are number of best practices of project management which helps the project manager to take benefits from repeatable standard functions, not considering about size Such as in Fig-1.

This figure shows the common best practices of project management which are applied by the project manager to complete a project in successful way. The signs of a successful project are knowing what needs to be done, a well understanding between the team members, no communication gap between all the stakeholders, a thorough planning and design, maximum output with minimum expectations, delivery on every milestone, stick to the plan, accommodate changes, testing on each step and having a broad vision [9].

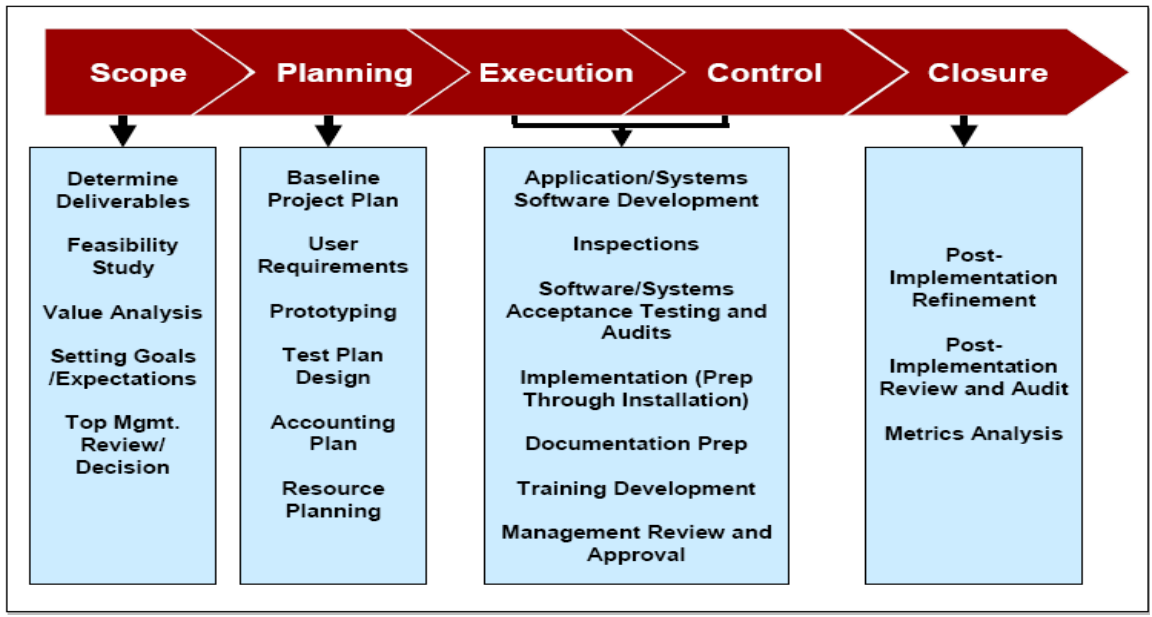

Fig-1. Best practices of Project management adopted by [6]

## 3 EXTREME PROGRAMMING (XP)

XP is most popular process model of AGILE methodology developed by Kent Beck and used to develop small size projects. XP projects start with a release planning phase, followed by several iterations, each of which concludes with user acceptance testing. When the product has enough features to satisfy users, the team terminates iteration and releases the software [7]. XP process model comprises on the number of best practices which are followed by developer to complete a project [10], besides this XP includes founding values which are Simplicity, communication, feedback and courage and all the best practices are coherent with these values. Kent Beck points out that the measurement in XP represents the basic management tools to get control on the project evolution [4].

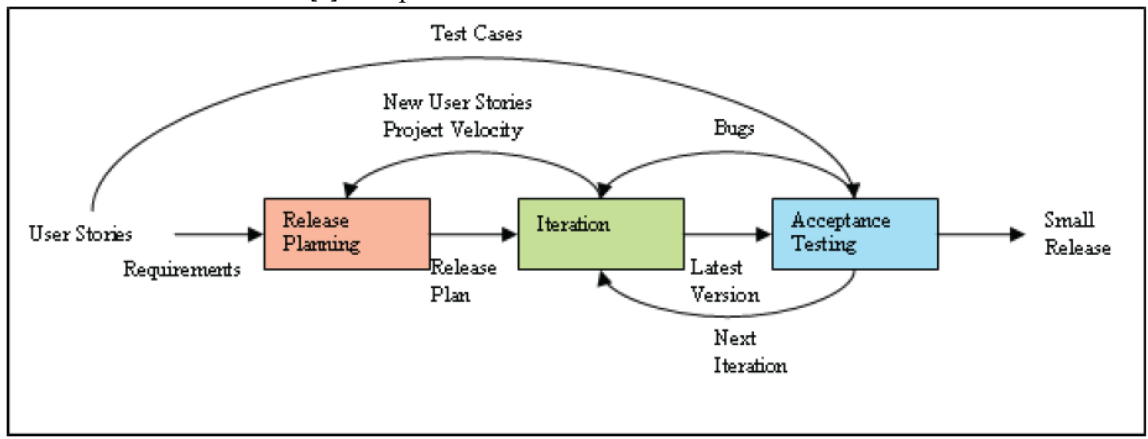

Fig-2. Overview of XP process adopted from [7]

## 4 MAPPING OF XP AND PROJECT MANAGEMENT BEST PRACTICES

Proper management of software project leads to developed successful, uncertain and quality software product and Project manager is the key resource of project management spectrum. In this paper authors mapped the best practices of XP with best activities of project management shown in Table-1. This table shows that how best practices of XP and project management can be mapped. In Table-1 five best practices of project management such as scope, planning, execution, controlling and closure are shown which are mapped with the best practices of XP. During development of XP project, project manager must know about its used practices and then follow the Table-1 to embed the project management activities.

**Mapping with Pair programming:**

Pair programming is the most common best practice of XP. According to this, two developers work in pairs, one as driver to write code and another as navigator to review code. In this practice, the project manager is responsible to select those roles for pairing who are interested and can work in collaboration. Besides this project manager will monitor the collaboration and other project issues relevant to pair programming.

**Mapping with On-Site customer:**

This practice shows the involvement of customer directly in development process. In this practice, the project manager is responsible to identify resources to gather consistent requirements and classification of resources i.e. whether off-line or on-line customers. Moreover, project manager monitor the implementation of requirements, relationship among requirements and satisfaction of requirements in consistent way.

**TABLE 1 MAPPING OF BEST PRACTICES OF XP AND PROJECT MANAGEMENT**

| S.No | Best Practices of XP | Project Management Activities | | | | |
|---|---|---|---|---|---|---|
| | | Scope | Planning | Execution | Control | Closure |
| 1 | Pair Programming | Supporting weightage | classify roles, Review process | collaboration | Collaboration | Collection of project issues |
| 2 | On site Customer | concern customers, on-line and off-line classification of customers, Sources of requirements | Requirements should be freeze, No repition and inconsistency of requirements, Dependency among requirements | Requirement's implementation, Removal of inconsistent and repeated requirement | satisfaction of each requirement, Impact on one requirement on another | Check requirements |
| 3 | Coding Standards | selected standard, supporting environment | Easily available, Match with environment | Easily applicable, Not effected on platform | Efficiency of adopting system, Causes no error | Causes no defects |
| 4 | Refactoring | Architecture of system will not effected, Impact on functionality | Abstraction level, Remove duplicate code | New code | Functionality, Abstraction Level | Defect rate |
| 5 | Collective code Ownership | Percentage of participation should be defined | Criteria for ownership | Should be applicable | Ownership propriety and its percentage | Should clear the ownership of all |
| 6 | Continuous Integration | Level of integration, Morale level | Day and time should be define, Risks identification | Integrate each build according to plan day and time | Risk monitoring, Morale level, Integration process | Defect rate due to integration process, Risks impacts |
| 7 | Release Planning | Define project's goals, On-site customer should defined | Requirement gathering through stories, Prioritize the stories | Satisfaction of stories | Priority of stories, Unique story, Related stories | Achieving project goals |
| 8 | Small Releases | Business value, Acknowledgment to Customer need | Number of Iterations, Prioritize the requirements | Each milestone should be unit tested, Release in sequence | Delivery of small release, Refinement process | Customer satisfaction |
| 9 | Sustainable pace | Technical roles | Identify technical roles and their duties, Maximum hour/day to work | Note work hour of each role | Monitor the time when role become tire and how it can sustain | Relaxation of developers, defect rate |
| 10 | Test Driven Development | Architecture, Functional | Write Automate unit test, Customer acceptance test | Test case for unit and user acceptance testing | long term benefits, Unit and user acceptance testing process | Achievement of benefits, Reduction in defects rate |
| 11 | Simple Design | Designing criteria | time management, documentation | Design Documentation | Time constraint, Design documentation | Effect of design documentation |
| 12 | Metaphor | New architecture, Adaptation | Availability of new architecture | Implementation Of new architecture | substitution of new architecture | Maintenance efforts |
| 13 | Stand up meeting | Allowable persons | Meeting date and time, Time management | Conduct meeting | Meeting process within time constraints | Requirement satisfaction |

**Mapping with Coding standards:**

This practice shows the adaptation of common coding conventions during whole development process. In this practice, the project manager is responsible to identify those coding conventions which are easily available, interpreted and adopted by all developers. Similarly project manager monitor portability, efficiency and defect rates after adopting new coding conventions.

**Mapping with Refactoring:**

This practice shows the improvement in existing code without changing the functionality of software. In this practice, the project manager is responsible to identify the code for refactoring, design new code and define the abstraction level for designing of new code. Moreover, project manager monitor the changes in functionality and defect rates which can arise due to designing of new code.

**Mapping with Collective code ownership:**

This practice shows the code will not be the proprietary of single person. In this practice, the project manager is responsible to define the criteria and percentage participation of all those developers who are involved in writing code. Moreover, project manager monitor demand of code ownership for all involved developers.

**Mapping with Continuous Integration:**

This practice shows the integration testing will be apply each time when a new build is developed and integrated into existing system. In this practice, the project manager is responsible to define the level of integration, date and time when a build will complete and integrated and involved risks. Moreover, project manager monitor the development and integration of new build according to schedule date and time, assessment of risk and defects rate due to integration process.

**Mapping with Release Planning:**

This practice shows the process to define user stories according to objectives and assign priority to each story either in initial stage or after completion of small release. In this practice, the project manager is responsible to monitor the uniqueness of user stories, prioritization of stories, dependency among stories, implementation of user stories and achievement of goals.

**Mapping with Small Releases:**

This practice shows the process to get feedback from user so early and focus on the effective growth of software in increments. In this practice, the manager is responsible to monitor the unit testing and delivery of each release, sequence of releases, feedback and satisfaction of customer after implementing new release.

**Mapping with Sustainable Pace:**

This practice shows that how a developer can work with efficiency after it had become tired. This is easy to define but harder to practice. Here the project manager is responsible to monitor the work of each role with in its job time and after duty time, note the efficiency of each role, relaxation of each role, effect of sustainable developers on project.

**Mapping with Test Driven Development:**

This practice shows automated unit and user acceptance testing process of each release. In this practice, the project manager is responsible to monitor the unit testing and acceptance testing of each release, achievement of long term benefits and defect rate after implementing.

**Mapping with Simple Design:**

This practice shows that design process of development will be simple and time will not waste in designing of documents. In this practice, the project manager is responsible to monitor the designing process of each architecture, wastage of extra time for documentation, and effect of no documentation.

**Mapping with Metaphor:**

This practice shows the process to substitute an existing architecture with new one. Here the project manager is responsible to monitor the availability of new architectures, adaptation and implementation of new architecture, substitution process, functionality of software and maintenance efforts.

**Mapping with Standup meeting:**

This practice shows the quick meeting process during collection of user stories. Here project manager is responsible to monitor the availability of person, time management, date and time for meeting and requirements satisfaction from user side.

## 5 CONCLUSION

A well defined structure of project management leads to develop a successful, quality and high productivity software. Often Projects fail due to poor planning and management process.

The effect of good project management is depend on the work of project manager who can managed that how activities can be performed in well-organized way. XP process model based on the AGILE methodology and it has a well defined project management structure. But in XP process model there is no proper guidelines for project manager to perform activities in effectively way. Here the purpose of authors has defined a process that how best practices of good project management can be mapped with best practices of XP. Moreover, this mapping technique is at its initial stage which can be enhanced later on after its successful implementation.

**Mr. Muhammad Javed**

I am an MS student in Institute of Computing and information technology, Gomal University D.I.Khan, Pakistan and I am also working here as lecturer on regular basis and teaching various subjects related to Software Engineering. I have got distinction throughout my academic carrier. I am doing specialization in the area of Cleanroom Software Engineering and RUP model.

**Mr. Shahid Hussain**

I did MS in Software Engineering from City University, Peshawar, Pakistan. I have got distinction throughout my academic carrier. He has done his research by introducing best practices in different software process models. I have introduced a new role communication model in RUP using pairing programming as best practice.. Recently, I am working as course chair cum Lecturer in Namal College, an associate college of University of Bradford. Moreover, I have published much research paper in different national/international journals and conferences such as MySec04, JDCTA, IJCSIS, NCICT,
ZABIST.